\documentclass{article}
\usepackage{amsmath, amsthm, amssymb}
 \usepackage{times}
\begin{document}
\numberwithin{equation}{section}
\def\bib#1{[{\ref{#1}}]}
\title{\bf Tomographic entropy and cosmology}
\author{{V.~I.~Man'ko${ }^{1,2}$, G.~Marmo${ }^{1,2}$ and C.~Stornaiolo ${ }^{1,2}$}\\
{\em $~^{1}$ Istituto Nazionale di Fisica Nucleare,}
{\em Sezione di Napoli,}\\
 {\em  Complesso Universitario di Monte S. Angelo}
 {\em Edificio N' }\\ {\em via Cinthia, 45 -- 80126 Napoli}\\
{\em $~^{2}$ Dipartimento di Scienze
Fisiche,}\\ {\em Universit\`{a} ``Federico II'' di Napoli,}\\
 {\em  Complesso Universitario di Monte S. Angelo}
 {\em Edificio N'  }\\ {\em via Cinthia, 45 -- 80126  Napoli}\\ }
\date{ }
\maketitle

\begin{abstract}
The probability representation of quantum mechanics including
propagators and tomograms of quantum states of the universe and its
application to quantum gravity and cosmology are reviewed. The
minisuperspaces  modeled by oscillator, free pointlike particle and
repulsive oscillator are considered. The notion of tomographic
entropy and its properties are used to find some inequalities for
the tomographic probability determining the quantum state of the
universe. The sense of the inequality as a lower bound for the
entropy is clarified.
\end{abstract}

\vskip 0.5truecm
\section{Introduction}\label{introduction}

Recently \cite{Manko:2003dp}
\cite{Man'ko:2004zj}\cite{Stornaiolo:2006kz}\cite{Stornaiolo:2006da}
a tomographic probability approach to describe the states of the
universe in quantum cosmology was suggested. In the framework of
this approach the quantum state of the universe is associated with
the standard positive probability distribution (function or
functional). The probability distribution contains the same
information on the universe quantum state that  the wave function of
the universe \cite{hawking1} \cite{whdw}\cite{hh} or the density
matrix of the universe \cite{page}, \cite{hawking2}. The latter can
be presented in different forms, e.g. in form of a Wigner function
\cite{wigner32} considered in  \cite{anton} in a cosmological
context. In fact the tomographic probability distribution describing
the state of the universe is a symbol of a density operator
\cite{marmops}\cite{marmoolga} and the tomographic symbols of the
operators realize one of the variants of the star-product
quantization scheme widely used \cite{fronsdal} to study the
relation of classical and quantum pictures \cite{marmo2}, which can
also be applied to study the relation of classical and quantum
descriptions of the universe in quantum cosmology. One of the
important ingredients of such descriptions is the evolution of the
state. In quantum mechanics such evolution is completely described
by means of a complex transition probability amplitude from an
initial state to a final one. This probability amplitude
(propagator) can be presented in the form of a Feynman path integral
containing the classical action. In quantum mechanics in the
probability representation using the tomographic approach the state
evolution can be associated with the standard transition
probability. It contains also information on the transition
probability amplitude related to the probability by integral
transform induced by the Radon transform relating the density matrix
(Wigner function) with the quantum tomographic probability
\cite{mancini}, \cite{lecture},  \cite{mendespl}, \cite{mendes2}.

In our previous work \cite{Manko:2003dp} we suggested to associate
the state of the universe in quantum cosmology with the tomographic
probability (or tomogram). The aim of our paper is to consider now
in the framework of the suggested probability representation of the
universe state in quantum cosmology also the cosmological dynamics
and to express this dynamics in terms of a positive transition
probability connecting initial and final tomograms of the universe.
Another goal of the work is to discuss the tomographic entropy of
the quantum state of the universe and a possible experimental
approach to observe the tomogram of the universe  at its present
stage and try to extract some information on the tomogram of the
initial state of the universe as well as to find some unilateral constraints
(inequalities). The idea of this attempt is based on the fact that
tomograms may describe the states of a classical system and the states of its
quantum counterpart. In this sense in the probability representation
of the quantum state there is not such a dramatic difference between
the classical and quantum pictures as the difference between wave
function (or density matrix) and classical probability distribution
(or trajectory) in the classical phase space. Due to this one can
try to study the cosmological dynamics namely in the tomographic
probability representation.

In order to illustrate the idea we will use the same simple
example of the universe description by means of the minisuperspace
discussed, e.g., in  \cite{hh}, \cite{Lemos:1995qu} . In these
minisuperspaces the quantum cosmological dynamics in operative
form is reduced to the dynamics of formal quantum systems
described by Hamiltonians of the types of oscillator, free motion
and free falling particles. In view of this one can apply the same
recently obtained results on description of  such systems by
tomographic probabilities  \cite{olga} to the cosmological
dynamics.

The paper is organized as follows. In the next section we will
review the cosmology in terms of a homogeneous (and isotropic)
metric with a time dependent parameter the expansion factor of the
universe. In section 3, we review the tomographic approach to
evolution of the quantum system. In section 4 we consider the
examples of the minisuperspace described by the reduced
Hamiltonians. In section 5 we study the tomographic entropy and its
evolution. Conclusions and perspectives are presented in section 6.

\section{The cosmological equations for an homogeneous and isotropic universe}\label{cosmology}

Let us recall briefly the equations for a classical homogeneous and
isotropic universe. It  is described by one of the following
metrics
\begin{equation}\label{metric}
    ds^{2}=-c^{2}dt^{2}+
    \frac{ a^{2}}{1-kr^{2}}\left(
    dr^{2}+r^{2}d\theta^{2}+r^{2}\sin^{2}\theta\right)d\phi^{2}
\end{equation}
where the parameter $k$ can  take positive, null, or negative
values related respectively to a closed universe, a flat universe
and an open universe respectively.

When the gravitational source is a perfect fluid, described by the
energy-momentum tensor, the Einstein equations with the metric
(\ref{metric}) may be given in the second order form

\begin{equation}\label{einstein1}
  \frac{\ddot{a}}{a}=-\frac{4\pi G}{3}(\rho+3P)
\end{equation}
which represents the dynamic equation and

\begin{equation}\label{einstein2}
    \frac{{\dot{a} }^{2}}{{a}^{2}}+\frac{k}{a^{2}}=\frac{8\pi G}{3}\rho
\end{equation}
which is a constraint, i.e. it defines the manifold of allowed
initial conditions. It takes a simple computation to show that
there are no secondary constraints. It constitutes an ``invariant
relation'', according to Levi-Civita.

From equations (\ref{einstein1}) and (\ref{einstein2}) the first
order equation
\begin{equation}\label{conservation}
     \dot{\rho}=-3\frac{\dot{a}}{a}\left(\rho+P\right)
\end{equation}
can be derived by taking the time derivative of (\ref{einstein2}).
It can be used alternatively in a system with equation
(\ref{einstein2}).

The system of equations (\ref{einstein1}) and (\ref{einstein2}) or
   (\ref{einstein2}) and (\ref{conservation}) are not complete, they  must be
completed by an equation of state $P=P(\rho)$ which is discussed in 
\cite{Stornaiolo:1994mw}. Usually a
linear equation of state is considered like  $P= (\gamma-1)\rho$ ($\gamma=1$ is the so-called matter
fluid, $\gamma=4/3$ is the radiation fluid and so on).

Equation (\ref{conservation}) together with an equation of state,
is important for our purpose  because it shows that the lefthand
side of equations (\ref{einstein1}) and (\ref{einstein2}) can be
expressed as a function of $a$ and represents a force in these
equations, if we treat them as equations for a ``point'' particle
as a result we have

 \begin{equation}\label{rhodia}
    \rho= \frac{\rho_{0}a_{0}^{3\gamma}}{a^{3\gamma}}
\end{equation}
when the equation of state is linear.

It is possible to derive the cosmological model from a point
particle Lagrangian, where the expansion factor $a$ takes the part
of the particle coordinate. Let us introduce the following
Lagrangian\cite{Man'ko:2004zj}

\begin{equation}\label{lagrangian}
   \mathcal{L}=3a\dot{a}^{2}-3ka +8\pi G \rho_{0}a_{0}^{3\gamma}a^{3(1-\gamma)}.
\end{equation}

The gravitational part is formally derived by substituting
directly metric (\ref{metric}) into the (field) general
relativistic action $\int\sqrt{-g}R$ and the material part is
obtained by putting a corresponding potential term $\Phi(a)=8\pi G
\rho_{0}a_{0}^{3\gamma}a^{3(1-\gamma)}$, in the case of a fluid
source.

 Equation (\ref{einstein1}) follows from the variational method applied to the
Lagrangian (\ref{lagrangian}).

From equation (\ref{lagrangian}) the conjugate momentum of $a$ is
\begin{equation}\label{momentum}
     p_{a}=\frac{\partial L}{\partial\dot{ a}}= 6 a \dot{a}.
\end{equation}

 Equation (\ref{einstein2}) is a constraint which is equivalent to the vanishing of the ``energy
 function'' $ E_{\mathcal{L}}$ associated to the Lagrangian
 \begin{equation}\label{energyfunction}
    E_{\mathcal{L}}=3a\dot{a}^{2}+ 3ka -8\pi G \rho_{0}a_{0}^{3\gamma}a^{3(1-\gamma)}.
\end{equation}

An alternative way to describe cosmology with a cosmological
fluid, with $\Lambda=0$,  was introduced firstly by Lemos
\cite{Lemos:1995qu}, Faraoni \cite{Faraoni:1999qu} and also in \cite{Man'ko:2004zj} and \cite{Lapchinsky:1977vb}.

They showed that equations (\ref{einstein1})  and
(\ref{einstein2}) can be transformed by means of a reparametrized time in equations similar to the
harmonic oscillator ones. By passing to the conformal time $\eta$,
defined by the relation
$$d\eta=\frac{dt}{a(t)},$$ and with the change of variables
 \begin{equation}\label{substitution}
w= {a}^{\chi}
\end{equation} where $$\chi=\frac{3}{2}\gamma-1 $$
equation (\ref{einstein1}) takes the form
\begin{equation}\label{harmonic}
    w''+k\chi^{2} w=0.
\end{equation}

Similarly it was shown in \cite{Stornaiolo:2006kz} that
cosmological equations with a perfect fluid and a cosmological
constant $\Lambda$
\begin{equation}\label{einsteinlambda}
  \frac{\ddot{a}}{a}=-\frac{4\pi G}{3}(\rho+3P)+\frac{2}{3}\Lambda \  \ \ \
  \
\mathrm{and}\ \ \ \ \ \
    \frac{{\dot{a} }^{2}}{{a}^{2}}+\frac{k}{a^{2}}=\frac{8\pi
    G}{3}\rho+\frac{\Lambda}{3}
\end{equation}
with the change of variables
 \begin{equation}\label{substitutionlambda}
z= {a}^{\sigma},
\end{equation}
where
$$\chi=\frac{3}{2}\,\gamma-1 \ \ \mathrm{and}\ \ \ \  \sigma=\,(1+\chi)^{-1}=\frac{2}{3\gamma}\,$$
are transformed into the  equation
\begin{equation}\label{harmoniclambda}
    \ddot{z} =\frac{\Lambda\gamma }{2\sigma}\,z+ \frac{k\chi}{\sigma}\frac{1}{z^{2\sigma-1}}.
    \end{equation}

    or
\begin{equation}\label{harmonic2}
    \ddot{z} =\frac{\Lambda}{3}\,z+k\left(1- \frac{2}{3\gamma}\right)
   z^{1-(4/3\gamma)} .
    \end{equation}
where the time variable is now the cosmic time and not the conformal
time as before.

Therefore a flat universe  ($k=0$) with a fluid and a cosmological
constant can be regarded again  as a harmonic oscillator (anti de
Sitter universe), a free particle (Einstein-de Sitter universe) and
a repulsive harmonic oscillator (de Sitter universe).

Similar considerations can be done also for  cosmological models
where the source is originated by a scalar field, which satisfies
the Klein-Gordon equation specialized to a homogeneous and isotropic
universe.  Also in this case, the evolution of universe can be
described by  equation (\ref{harmonic}). In \cite{gousheh} there are
other examples in which the cosmological models with a scalar field
can be described by equations similar to (\ref{harmonic}).

\section{Evolution in minisuperspace in the framework of
tomographic probability representation}\label{evolution}

We will discuss below the evolution of a universe in the framework
of the minisuperspace model discussed in the previous section.
Thus the state of the universe is described by a wave function
$\Psi(x,t)$. This wave function evolves in time from its initial
value $\Psi(x,t_{0})$ and this evolution can be described by a
propagator $G(x,x',t,t_{0})$
\begin{equation}\label{propagatorevol}
\Psi(x,t)=\int G(x,x',t,t_{0})\Psi(x',t_{0})dx'.
\end{equation}
The propagator can be obtained using path integration over
classical trajectories of the exponential of the classical action
$S$
\begin{equation}\label{propagatoraction}
G(x,x',t,t_{0})= \int D[x(t)] e^{\frac{iS[x(t)]}{\hbar}}.
\end{equation}

In our previous work \cite{Manko:2003dp} we discussed the
properties of the new representation (tomographic probability
representation) of the quantum states of the universe.

In this representation (which we discuss below in the framework of
a minisuperspace model) the wave function of the universe
$\Psi(x,t)$ or the density matrix of the universe

\begin{equation}\label{rho}
\rho(x,x',t)=\Psi(x,t)\Psi^{*}(x',t)
\end{equation}

can be mapped onto the standard positive distribution ${\cal W}(X,
\mu,\nu, t)$ of the random variable $X$ depending on the two real
extra parameters $\mu$ and $\nu$ and the time $t$. The map is
given by the formula (we take $\hbar=1)$
\begin{equation}\label{probdistribution}
    {\cal W}(X, \mu,\nu, t)= \frac{1}{2\pi |\nu|}\int\rho(y,y',t)
    e^{i\frac{\mu(y^{2}-{y'}^{2})}{2\nu}-i\frac{X}{\nu}(y-y')}dy'
    dy.
\end{equation}

In fact, equation (\ref{probdistribution}) is the fractional
Fourier transform \cite{margarita} \cite{marg} of the density
matrix. The map has inverse and the density matrix can be
expressed in terms of the tomographic probability representation
as follows

\begin{equation}\label{densitytomograph}
    \rho(x,x',t)=\frac{1}{2\pi} \int
    \mathcal{W}(Y,\mu,x-x',t)\,
    e^{i\left(Y-\frac{\mu}{\nu}(x+x')\right)}dYd\mu.
\end{equation}
The expression (\ref{probdistribution}) can be given an affine
invariant form \cite{mendes3}
\begin{equation}\label{invariantform}
  {\cal W}(X, \mu,\nu, t)= \left\langle \delta(X-\mu\hat{q}-\nu\hat{p})\right\rangle
\end{equation}
Here $\langle\ \rangle$ means trace with the density operator
$\hat{\rho} (t)$ of the universe state, $\hat{q}$ and $\hat{p}$
are the operators of position (universe expansion factor) and the
conjugate moment respectively. From equation (\ref{invariantform})
some properties of the tomogram ${\cal W}(X, \mu,\nu, t)$ are
easily extracted. First, the universe tomogram is a normalized
probability distribution, i.e.
\begin{equation}\label{normalized}
    \int{\cal W}(X,
\mu,\nu, t) dX=1
\end{equation}
if the universe density operator is normalized (i.e. $Tr
\hat{\rho}(t)=1$). Second, the tomogram of the universe state has
the homogeneity property \cite{rosapl}
\begin{equation}\label{homogeneity}
    {\cal W}(\lambda X,
\lambda\mu,\lambda\nu, t)= \frac{1}{|\lambda|} {\cal W}(X,
\mu,\nu, t)
\end{equation}
The tomogram can be related with  such quasidistribution as the
Wigner function $W(q,p,t)$ \cite{wigner32} used in the phase space
representation of the universe states in \cite{anton}.

The relation reads

\begin{equation}\label{wignertomo}
  {\cal W}(X,\mu,\nu,t)= \int W(q,p,t)\delta(X-\mu q - \nu p)\frac{dqdp}{2\pi}
\end{equation}

which is the standard Radon transform of the Wigner function. The
physical meaning of the tomogram $ {\cal W}(X,\mu,\nu,t)$ is the
following. One has in the phase space the line
\begin{equation}\label{line}
    X=\mu q + \nu p
\end{equation}
which is given by equating to zero of the delta-function argument in
equation (\ref{wignertomo}). The real parameters $\mu$ and $\nu$
can be given in the form
\begin{equation}\label{parameters}
    \mu=s\cos\theta~~~~~~~~~~~~~~~~~\nu=s^{-1}\sin\theta.
\end{equation}
Here $s$ is a real squeezing parameter and $\theta$ is a rotation
angle. Then the variable $X$ is identical to the position measured
in the new reference frame in the universe phase-space. The new
reference frame has new scaled axis $sq$ and $s^{-1}p$ and after
the scaling the axis are rotated by an angle $\theta$. Thus the
tomogram implies the probability distribution of the random
position $X$ measured in the new (scaled and rotated) reference
frame in the phase-space. The remarkable property of the
tomographic probability distribution is that it is  a fair positive
probability distribution and it contains a complete information of
the universe state contained in  the  density operator
$\hat{\rho}(t)$ which can be expressed in terms of the tomogram as
\cite{d'ariano}
\begin{equation}\label{operatorrho}
    \hat{\rho}(t)=\frac{1}{2\pi}\int \mathcal{W}(X,\mu,\nu,t)e^{i(X-\mu\hat{q}-\nu\hat{p})}
dX d\mu d\nu
\end{equation}
Formulae (\ref{invariantform}) and (\ref{operatorrho}) can be
treated with the tomographic star-product quantization schemes
\cite{marmoolga} used to map the universe quantum observables
(operators) onto functions (tomographic symbols) on a manifold
$(X,\mu,\nu)$. The tomographic map can be used not only for the
description of the universe state by probability distributions,
but also to describe the evolution of the universe (quantum
transitions) by means of the standard real positive transition
probabilities (alternative to the complex transition probability
amplitudes). The transition probability
$$\Pi(X,\mu,\nu,t,X',\mu',\nu',t_{0})$$ is the propagator  expressed in tomographic representation, it
gives the tomogram of the universe  ${\cal W}(X,\mu,\nu,t)$, if
the tomogram at the initial time $t_{0}$ is known, in the form
\begin{equation}\label{proptomo}
     {\cal W}(X,\mu,\nu,t)= \int \Pi(X,\mu,\nu,t,X',\mu',\nu',t_{0}){\cal
     W}(X',\mu',\nu',t_{0})dX' d\mu'd\nu'.
\end{equation}
The positive transition probability describing the evolution of
the universe has the obvious nonlinear properties used in
classical probability theory, namely
$$ \Pi(X_{3},\mu_{3},\nu_{3},t_{3},X_{1},\mu_{1},\nu_{1},t_{1})=\int
 \Pi(X_{3},\mu_{3},\nu_{3},t_{3},X_{2},\mu_{2},
\nu_{2},t_{2})$$
\begin{equation}\label{postransprob}
~~~~~~~~~~~~~~~\times\Pi(X_{2},\mu_{2},\nu_{2},t_{2},X_{1},\mu_{1},\nu_{1},t_{1})\,dX_{2}\,d\mu_{2}\,d\nu_{2}.
\end{equation}
They follow from the associativity property of the evolution maps.
This nonlinear relation is the tomographic version of the nonlinear relation of
the complex quantum propagators of the universe wave function
 \begin{equation}\label{complexnonlinear}
 G(x_{3},x_{1},t_{3},t_{1})=\int G(x_{3},x_{2},t_{3},t_{2})
 G(x_{2},x_{1},t_{2},t_{1})dx_{2}.
\end{equation}
Both relations (\ref{postransprob}) and (\ref{complexnonlinear})
imply that the state of the universe evolves from the initial one
  to the final one   through all
intermediate states. The remarkable fact is that this quantum
evolution of the universe state can be associated with a
standard positive transition probabilities like in classical
dynamics. This is connected with the existence of the invertible
relations of the tomographic and quantum propagators
\cite{mendes2}\cite{olga} . If one denotes
\begin{equation}\label{kappa}
    K(X,X',Y,Y',t)=G(X,Y,t)G^{*}(X',Y',t),
\end{equation}
then the quantum propagator may be given the following form
$$ K(X,X',Y,Y',t)= \frac{1}{(2\pi)^{2}}\int \frac{1}{|Y'|}
\exp\left\{i\left(Y-\mu \frac{(X+X')}{2}
\right)-i\frac{Z-Z'}{\nu'}Y'\right.$$

\begin{equation}\label{kappa1}
\left. + i
\frac{Z^{2}-{Z'}^{2}}{2\nu'}\mu'\right\}\Pi(Y,\mu,X-X',0,X',\mu',\nu',t)d\mu\,d\mu'\,dY\,dY'\,d\nu'.
\end{equation}

This relation can be reversed. Thus the propagator for the
tomographic probability can be expressed in terms of the Green
function $G(x,y,t)$ as follows (we take $t_{0}=0$)
$$\Pi(X,\mu,\nu,X',\mu',\nu',t)=\frac{1}{4\pi}\int
    k^{2}G(a+\frac{k\nu}{2}, y,t)G^{*}(a-\frac{k\nu}{2},
    z,t)\delta(y-z-k\nu') $$

\begin{equation}\label{inverse}
   \times \exp\left[ik(X'-X+\mu q-\mu'\frac{y+z}{2})\right]dk dy dz
   dq.
\end{equation}
The relation can be used to express the tomographic propagation in
terms of the Feynmann path integral using the formula for the
quantum propagator (\ref{propagatoraction}) where the classical
action is involved. It means that the positive transition
probabilities (\ref{inverse}) can be reexpressed in terms of the
double path integral (with four extra usual integrations).

The discussed relations demonstrate that the  quantum universe
evolution can be described completely using only positive
transition probabilities.

Standard complex transition probability amplitudes (and Feynman
path integral) can be reconstructed using this transition
probability by means of equation (\ref{kappa1}.)

\section{Evolution of the universe in the oscillator model
framework}\label{oscillator}

As we have shown the equation for the universe evolution in the
conformal time picture (\ref{harmonic}) can be cast in  the form
of an oscillator equation. The oscillator has the frequency
$\omega^{2}=\pm k\chi^{2}$.

For $k=0$ one has the model of free motion. For $k<0$ one has the
model of a inverted oscillator and for $k>0$ one has the
standard oscillator as solution of the equation (\ref{harmonic}).
We assume below that the quantum behavior of the
universe in the framework of the considered minisuperspace model
is described by the quantum behavior of the oscillator as derived in the previous sections. 
Though the connection (\ref{substitution}) of the expansion factor $a(\eta)$
with the classical observable $w$ which obeys to oscillator motion
provides constraints on the   ranging domain of this variable, we
assume in the quantum picture of the variable to lie on the real
line $R$. In such approach we apply the tomographic probability
representation, developed in the last section, to quantum states
of the universe in the framework of the oscillator model. We will
denote  in the quantum description the variable as $q$
 ($w\rightarrow q$)and the conformal time as $t$\  ($\eta\rightarrow
t$). Thus the tomographic probability $\mathcal{W}(X,\mu,\nu,t)$of
the universe state obeys the evolution equation
\cite{Manko:2003dp} for the potential energy $V(q)$ in the form
$$\frac{\partial\mathcal{W}(X,\mu,\nu,t)}{\partial t}-
     \mu\frac{\partial\mathcal{W}(X,\mu,\nu,t)}{\partial \nu}+
     i\left[V\left(-\left(\frac{\partial}{\partial X}\right)^{-1}\frac{\partial}{\partial\mu}
     -\frac{i\nu}{2}\frac{\partial}{\partial X}\right)\right.$$
\begin{equation}\label{tomoequation1}
\left. - V\left(-\left(\frac{\partial}{\partial
X}\right)^{-1}\frac{\partial}{\partial\mu}
     +\frac{i\nu}{2}\frac{\partial}{\partial
     X}\right)\right]\mathcal{W}(X,\mu,\nu,t)=0,
\end{equation}
where the operator $(\partial/\partial X)^{-1} $ is defined by the
relation
\begin{equation}\label{relation}
\left(\frac{\partial}{\partial X}\right)^{-1}\int
f(y)e^{iyX}dy=\int \frac{f(y)}{(iy)}e^{iyX}dy.
\end{equation}
The propagator of this equation $\Pi(X,\mu,\nu,t,X',\mu',\nu')$
satisfies equation (\ref{tomoequation1}) with the extra term
$$\frac{\partial\Pi}{\partial t}-
     \mu\frac{\partial\Pi}{\partial \nu}+
     i\left[V\left(-\left(\frac{\partial}{\partial X}\right)^{-1}\frac{\partial}{\partial\mu}
     -\frac{i\nu}{2}\frac{\partial}{\partial X}\right)\right.$$
\begin{equation}\label{tomoequation2}
\left. - V\left(-\left(\frac{\partial}{\partial
X}\right)^{-1}\frac{\partial}{\partial\mu}
     +\frac{i\nu}{2}\frac{\partial}{\partial
     X}\right)\right]\Pi=\delta(\mu-\mu')\delta(\nu-\nu')\delta(X-X')\delta(t),
\end{equation}
For the considered model the general equation for the universe
tomogram evolution takes the simple form of a first order
differential equation
\begin{equation}\label{tomoequation3}
\frac{\partial\mathcal{W}}{\partial t}-
     \mu\frac{\partial\mathcal{W} }{\partial \nu}
      + \omega^{2}\nu\frac{\partial\mathcal{W} }{\partial \mu} =0.
\end{equation}
Analogously for the propagator of the tomographic equation for the
universe in the framework of the oscillator model one has
\begin{equation}\label{tomoequation4}
\frac{\partial\Pi }{\partial t}-
     \mu\frac{\partial\Pi }{\partial \nu}
      + \omega^{2}\nu\frac{\partial\Pi }{\partial \mu} =\delta(\mu-\mu')\delta(\nu-\nu')\delta(X-X')\delta(t).
\end{equation}

A solution to this equation can be found to be in the case $k>0$
$$ \Pi^{osc.}(X,\mu,\nu,t,X',\mu',\nu')= \delta(X-X') \delta(\mu'-\mu\cos\,
\omega t+ \omega\nu\sin \omega t)$$
\begin{equation}\label{solutionk1}
\times \delta\left(\nu'-\nu\cos\, \omega t -
\frac{\mu}{\omega}\sin\, \omega t\right).
\end{equation}

In the limit $k=0$ (free motion) the equation for the tomogram
(\ref{tomoequation3}) becomes
\begin{equation}\label{tomoequation5}
\frac{\partial\mathcal{W}(X,\mu,\nu,t)}{\partial t}-
     \mu\frac{\partial\mathcal{W}(X,\mu,\nu,t)}{\partial \nu} =0.
\end{equation}

The corresponding propagator solution  reads

\begin{equation}\label{solutionk2}
  \Pi^{free}(X,\mu,\nu,t,X',\mu',\nu')= \delta(X-X')
\delta(\mu'-\mu) \delta(\nu'-\nu -\mu t).
\end{equation}
Finally for the case $k<0$   the propagator has the form
corresponding to a repulsive oscillator
$$ \Pi^{rep.}(X,\mu,\nu,t,X',\mu',\nu')= \delta(X-X') \delta(\mu'-\mu\cosh\,
\omega t -  \omega\nu\sinh \omega t)$$
\begin{equation}\label{solutionk3}
\times \delta\left(\nu'-\nu\cosh\, \omega t -
\frac{\mu}{\omega}\sinh\, \omega t\right).
\end{equation}
Thus we got the dynamics of the universe given by the transition
probabilities $\Pi^{osc.}$, $\Pi^{free}$ and $\Pi^{rep.}$ for the
three cases $k>0$, $k=0$ and $k<0$ respectively. One can see that
this dynamics is compatible with the dynamics calculated in the
standard representation of the complex Green function (quantum
propagator). For  $k=1$ the form of the Green function reads
\begin{equation}\label{Greenk1}
     G^{osc.}(X,X',t)=\sqrt{\frac{\omega}{2\pi i \sin \omega t}}
     \exp\left\{\frac{i\omega}{2}\left[\cot \omega t
     \left(X^{2}+{X'}^{2}\right)-\frac{2XX'}{\sin \omega t}\right]\right\}
\end{equation}
For the case of the free motion model the Green function can be
obtained by the limit $\omega\to 0$ in this expression and one has
\begin{equation}\label{Greenk2}
     G^{free}(X,X',t)=\sqrt{\frac{1}{2\pi i t}}
     \exp\left[i
     \frac{\left(X-{X'}\right)^{2}}{2t}\right]
\end{equation}
and for the repulsive oscillator model one has
\begin{equation}\label{Greenk3}
     G^{rep.}(X,X',t)=\sqrt{\frac{\omega}{2\pi i \sinh \omega t}}
     \exp\left\{\frac{i\omega}{2}\left[\coth \omega t
     \left(X^{2}+{X'}^{2}\right)-\frac{2XX'}{\sinh \omega
     t}\right]\right\}.
\end{equation}
All these three universe cases can be discussed using the Green
function in terms of the Feynmann path integral.

Thus the expression (\ref{Greenk1}) is given  by the formula
\begin{equation}\label{feynmanngreen}
     G(X,X't)=\int e^{i\int_{0}^{t}\left[\frac{\dot{x}^{2}(t)}{2}-
     \frac{\omega^{2}x^{2}(t)}{2}\right]dt}D[x(t)]
\end{equation}

The integral in the exponent of the path integral provides the
classical action for the oscillator
\begin{equation}\label{classical}
  S^{cl.}(X,X',t)= \int_{0}^{t}\left[\frac{\dot{x}^{2}(t)}{2}-
     \frac{\omega^{2}x^{2}(t)}{2}\right]dt
\end{equation}
where the trajectories start at $t=0$ at $X'$ and end at time $t$
at the point $X$. The classical action satisfies the
Hamilton-Jacobi equation

\begin{equation}\label{hamilton}
    \frac{\partial S^{cl.}(q,q',t)}{\partial t} +
    \mathcal{H}\left(q,p=- \frac{\partial S^{cl.}(q,q',t)}{\partial q}\right)=0
\end{equation}
where $\mathcal{H}$ is the Hamiltonian
\begin{equation}\label{hamiltonian}
 \mathcal{H}=\frac{p^{2}}{2}+\frac{\omega^{2}q^{2} }{2}
\end{equation}
For the free motion  model one has

\begin{equation}\label{freen}
  G^{free}(X,X',t)=\int
  e^{i\int_{0}^{t}\frac{\dot{x}^{2}(t)}{2}dt}D[x(t)].
\end{equation}
The path integral is integrated and the result (\ref{Greenk2})
contains in the exponent term the classical action
\begin{equation}\label{action}
     S^{(f)}(X,X',t)=\frac{(X-X')^{2}}{2t}
\end{equation}
which is solution of the Hamilton-Jacobi equation with the
Hamiltonian
\begin{equation}\label{freehamiltonian}
     \mathcal{H}=\frac{p^{2}}{2}.
\end{equation}
For the repulsive model one has the same structure of path
integral and the result of path integration is expressed in terms
of the classical action
\begin{equation}\label{classicalrepulsive}
    S^{rep.}(X,X',t)= \frac{ \omega}{2}\left[\coth \omega t
     \left(X^{2}+{X'}^{2}\right)-\frac{2XX'}{\sinh \omega
     t}\right],
\end{equation}
which is solution of the Hamilton-Jacobi equation with the
Hamiltonian
\begin{equation}\label{hrep}
    \mathcal{H}^{rep.}=\frac{p^{2}}{2}-\frac{\omega^{2}q^{2} }{2}.
\end{equation}
All the obtained propagators complex Green
functions or path integrals are related with the propagators in
probability representation by means of equations
(\ref{postransprob}) and (\ref{complexnonlinear}).

Thus the universe evolution can be described in the oscillator
model of minisuperspace  for $k>0$, $k=0$ and $k<0$ by means of
the standard transition probabilities expressed as propagators
$\Pi^{osc.}$, $\Pi^{free}$ and $\Pi^{rep.}$ respectively.
\section{Entropy in cosmological models}

With any symplectic tomogram ${\cal W}(X,\mu,\nu,t)$ one can
associate the tomographic entropy \cite{marg} \cite{lecture}
\begin{equation}\label{tomoentropy}
     \mathcal{S}(\mu,\nu,t)= -\int {\cal W}(X,\mu,\nu,t) \ln {\cal
     W}(X,\mu,\nu,t) dX\, .
\end{equation}
The standard Von Neumann entropy of the quantum state $ S_{VN}=
-Tr\hat{\rho}\ln\hat{\rho}$ is minimum in the tomographic entropy
for a finite Hilbert space. But the Von Neumann entropy does not
distinguish any pure state because it is equal to zero for arbitrary
$\hat{\rho}_{\psi}=|\psi><\psi|$. So we hope that the tomographic
entropy (\ref{tomoentropy}) may characterize the chaoticity
properties of different states of the universe at least for the
cases of minisuperspace models. For example for the oscillator
model in the case of a Fock state $|n><n|$ the entropy reads (at a
given time moment)

  $$\mathcal{ S}_{n}(\mu,\nu)= -\int \left[ \frac{1}{2^{n}}\frac{1}{n!}\frac{\exp(-X^{2}/(\mu^{2}+\nu^{2}))}{\sqrt{\pi}(\mu^{2}+\nu^{2})}
   H^{2}_{n}\left(\frac{X}{\sqrt{(\mu^{2}+\nu^{2})}}\right)\right.$$
\begin{equation}\label{entropyoscillator}
   \left. \times\ln
   \left(\frac{1}{2^{n}}\frac{1}{n!}\frac{\exp(-X^{2}/(\mu^{2}+\nu^{2}))}{\sqrt{\pi}(\mu^{2}+\nu^{2})}H^{2}_{n}\left( X \right)\right)\right]dX\,
   .
\end{equation}
For any state of the universe with generic Gaussian Wigner function
the tomogram is also a generic Gaussian distribution and it evolves
with the time evolution of the universe.

The evolution for the minisuperspace oscillator model is governed by
a classical evolution equation. Due to this we hope to compare the
entropy values \cite{Hartle:1997hw} with the numbers obtained from symplectic
tomographic entropy. Also we hope to find a  relation of the
Bekenstein entropy bound with the introduced entropy of the quantum
universe state.

The properties of the tomographic entropy are studied in \cite{renato1}\cite{renato2}.

Since the tomogram of a quantum state satisfies the homogeneity
condition (\ref{homogeneity}) the tomographic entropy has the
property \cite{renato2}
\begin{equation}\label{1}
     S(\sqrt{\mu^{2}+\nu^{2}}\cos \theta,\sqrt{\mu^{2}+\nu^{2}}\sin \theta, t
     )-\frac{1}{2}\sqrt{\mu^{2}+\nu^{2}}=f(\theta,t)
\end{equation}
We used the polar coordinates
\begin{eqnarray}
\nonumber   \mu &=& \sqrt{\mu^{2}+\nu^{2}}\cos \theta \\
  \nu &=& \sqrt{\mu^{2}+\nu^{2}}\sin \theta.
\end{eqnarray}
Also, the particular values of the entropy
\begin{equation}\label{2}
    S(1,0,t)=-\int\rho(x,x,t)ln[\rho(x,x,t)]dx
\end{equation}
and
\begin{equation}\label{e4}
    S(0,1,t)=-\int\rho(p,p,t)ln[\rho(p,p,t)]dp
\end{equation}
are the Shannon entropies \cite{shannon}\cite{renyi} connected with the probability
distribution densities in position and momentum respectively.

It is known \cite{bialynicki} \cite{183} that the entropies satisfy the following
inequality
\begin{equation}\label{e5}
     S(1,0,t)+S(0,1,t)\geq \ln(\pi e).
\end{equation}
This inequality was extended \cite{renato1}\cite{renato2} to give the inequality for
the tomogram of the quantum state
\begin{equation}\label{e6}
   \int\left[  \mathcal{W}(X, \theta, t)\ln\mathcal{W}(X, \theta, t)+
   \mathcal{W}(X, \theta + \frac{\pi}{2}, t)\ln\mathcal{W}(X, \theta + \frac{\pi}{2},
   t)\right]dX+ \ln(\pi e)\leq 0.
\end{equation}
Here
\begin{equation}\label{e7}
     \mathcal{W}(X, \theta,
t)=\mathcal{W}(X,\mu=\cos\theta,\nu=\sin \theta, t) .
\end{equation}
The inequality (\ref{e6}) can be used to check whether the tomogram
$\mathcal{W}(X,\mu,\nu, t)$ satisfies the quantum constraints or
not. One can check that the entropy of an excited state of the
universe in the framework of the oscillator minisuperspace model
given by eq. (\ref{entropyoscillator}) satisfies the above
inequality. In this case since the angle $\theta$ disappears from
the lefthand side  of eq.(\ref{entropyoscillator}) the inequality
takes the form of an integral inequality for the Hermite polynomial.
\begin{equation}\label{e8}
   \frac{1}{2^{n}\sqrt{\pi}}\frac{1}{n!}\int \left[ e^{-X^{2}}
   H^{2}_{n}(x)
\ln   \frac{1}{2^{n}}\frac{1}{n!\sqrt{\pi}} e^{-X^{2}}H^{2}_{n}(
X)\right]dX\, + \frac{1}{2}\ln (\pi e)\leq 0
   .
\end{equation}

In the case of the Gaussian coherent states the inequality is
saturated and becomes equality. For example, the ground state with
wave function
\begin{equation}\label{e9}
    \psi_{0}=\frac{e^{-\frac{q^{2}}{2}}}{\sqrt[4]{\pi}}
\end{equation}
the tomogram reads $\mathcal{W}(X, \cos \theta, \sin
\theta)=e^{-X^{2}}/\sqrt{\pi}$ and
\begin{equation}\label{e10}
    \frac{1}{\sqrt{\pi}}\int e^{-X^{2}}\ln\frac{e^{-X^{2}}}{\sqrt{\pi}}+ \frac{1}{2}\ln (\pi
    e)=0.
\end{equation}
The discussed inequalities called entropic uncertainty  principle
are connected with the Heisenberg uncertainty relations. For Gaussian coherent
states (or ground state) the entropic inequality (\ref{e6})is
equivalent to
\begin{equation}\label{e11}
     (\delta q)^{2}(\delta p)^{2} \geq \frac{1}{4}
\end{equation}
where $(\delta q)^{2}$ and  $(\delta q)^{2}(\delta p)^{2}$ are the
dispersion of the position and the momentum when are both equal to
$1/2$.

Calculating the entropies $S(1,0)$ and $S(0,1)$ for Gaussians with
these dispersions one can find that the entropic inequalities in
this case provide the inequality (\ref{e11}). Thus the quantum
Heisenberg  uncertainty relation can be cast for the universe in the
framework of minisuperspace model as constraint for the universe
state tomogram. Thus, if one can extract some experimental data on
the universe state tomogram the tomographic entropy can be used to
control the compatibility of the quantum gravity constraints with
the observable data. It is worthy to note that only the quantum
tomogram must satisfy the discussed inequality. On the other side
the classical state may not to satisfy it. The inequality (\ref{e6})
can be written in another form. In fact, in the lefthand side  of this
inequality we have the periodic function of the angle $\theta$.
Expanding this function into Fourier series we get
\begin{equation}\label{e12}
   \int\left[  \mathcal{W}(X, \theta, t)\ln\mathcal{W}(X, \theta, t)+
   \mathcal{W}(X, \theta + \frac{\pi}{2}, t)\ln\mathcal{W}(X, \theta + \frac{\pi}{2},
   t)\right]dX= \sum c_{m}(t)e^{im\theta}.
\end{equation}
Averaging the function over the angles $\theta$ we obtain the
inequality
\begin{equation}\label{e13}
    c_{0}(t)+ln(\pi e) \leq 0.
\end{equation}
Here the constant contribution to the Fourier series depending on
time is the quantum state characteristic which is the functional of
the quantum tomogram of the universe. We discussed the one mode
case.

The number $\ln(\pi e)=2.14$ can be interpreted as the ``entropy''
of the vacuum state of the mode. In fact, if one takes the ground
state of the harmonic oscillator,  the Shannon entropy associated to
this state using the probability distribution in position equals the
Shannon entropy associated to the distribution in momentum. Each of
these equal entropies read $S_{x}=S_{p}=1/2+\ln(\pi)$.

So we can associate the positive minimal entropy  $\ln(\pi e)$  to
the vacuum state analogously to the minimal ground state energy.
This energy creates the notion of Casimir energy for many modes.

We can suggest that this is the analog of the vacuum dimensionless
entropy associated to the $\ln(\pi e)$ term for each mode.

Some important inequalities for entropy related to gravity have been
discovered by Bekenstein \cite{Bekenstein:1980jp} and by Verlinde
\cite{verlinde}, The constant $c_{0}$ is the mean value of the sum
of Shannon entropies of the probability distributions for two
conjugate variables (position and momentum). In classical mechanics
there is no correlations between these two entropies. In quantum
approach there appeared  such kind of correlation expressed in terms
of the inequality (\ref{e13}). For the case of several degrees of
freedom the bound of the inequality takes an integer factor.
becoming $N\ln \pi e$ where $N$ is the number of degrees of freedom.
Thus the entropy bound is ``quantized'' depending on the number of
degrees of freedom corresponding to the minisuperspace model.

 A straightforward calculation shows that we can also produce the
entropy evolution by substituting equation (\ref{proptomo}) into
equation (\ref{tomoentropy}). We find
  $$ \mathcal{S}(\mu,\nu,t)= -\int \int \Pi(X,\mu,\nu,t,X',\mu',\nu',t_{0}){\cal
     W}(X',\mu',\nu',t_{0})dX' d\mu'd\nu'$$
\begin{equation}\label{tomoentropy2}
    \times \ln \int \Pi(X,\mu,\nu,t,X',\mu',\nu',t_{0}){\cal
     W}(X',\mu',\nu',t_{0})dX' d\mu'd\nu' dX\, .
\end{equation}
which can be compared with the initial entropy
\begin{equation}\label{tomoentropy3}
     \mathcal{S}(\mu,\nu,t_{0})= -\int {\cal W}(X,\mu,\nu,t_{0}) \ln {\cal
     W}(X,\mu,\nu,t_{0}) dX\, .
\end{equation}

Thus the tomographic transition probability (propagator) determines
the evolution of the tomographic entropy. The form of the propagator
is compatible with the constraints (inequalities) which must be
fulfilled for the tomographic entropy and for the tomogram.

\section{Conclusions}\label{conclusions}

To conclude we discuss the main results of the work. In addition
to what suggested in \cite{Manko:2003dp}, the probability
representation of the universe quantum states for which the states
(e.g. of the universe in a minisuperspace model) are described by
the standard positive probability distribution, we introduce the
description of the universe dynamics by means of standard
transition probabilities.

The transition probabilities are determined as propagators
(integral kernels) providing the evolution of the universe
tomograms. It is shown that there is a relation of the standard
propagator determining the quantum evolution of the universe wave
function to the tomographic propagator. This relation permits to
reconstruct the complex propagator  for the wave function in terms
of the positive propagator for the universe tomogram. Also, one
can express the propagator for the tomogram in terms of the
propagator for the wave function of the universe.

These relations between the propagators mean that the Feynmann
path integral formulation or the universe properties (in quantum
gravity) contains the same information that the probability
representation of the quantum states of the universe including the
universe quantum evolution. As the simplest example of the
suggested transition probability picture, we considered the
minisuperspace model for which classical and quantum evolution is
described by the harmonic vibrations  in conformal time
\cite{Lemos:1995qu}, \cite{Faraoni:1999qu}, \cite{gousheh},
\cite{Lapchinsky:1977vb}.
 The specific property of this minisuperspace model is that the
tomographic propagators for both classical and quantum universe
tomograms coincide. This fact provides some possibility to connect
observations related to today classical epoch of the universe  and
its purely initial quantum state. In the framework of the
suggested approach (and in the framework of the considered
oscillator model), the universe evolution can be studied using
specific properties of the tomographic propagator. If one
considers  tomograms and their evolution in classical mechanics
\cite{lecture}\cite{mendes2} the specific property of the linear
systems (e.g. oscillator model) is that the tomographic
propagators in quantum and classical domains are in one-to-one
correspondence and are given in the same carrier space, therefore
we may say that in this picture the difference of the quantum and classical
evolution is in the initial conditions,  and relies on the fact that the product of functionals
on the ``wave functions'' are multiplied pointwise in the classical picture and non-locally in the quantum picture. 
This non-local product contains all the information of the indetermination relations and consequent 
constraints on the allowed tomograms.

They must satisfy uncertainty relations. The
choice of initial conditions (initial tomogram of the universe) in
correspondence with the uncertainty relation provides a
possibility to avoid the singularity of the metric, which is
unavoidable in the classical picture. But the following evolution
of the universe coded by the tomographic propagator is the same
(for the oscillator model).

Due to this result, one can extract from the present observational
classical data conclusions over the cosmological evolution. Evolving
backwards in time the present situation by means of the ``true''
quantum or the classical propagators, we may find discrepancies
between the initial conditions at minus infinity. Using the notion
of tomographic entropy we also established the specific constraints
for the quantum states of the universe expressed in terms of  an
inequality. This inequality must be satisfied in the quantum
mechanical approach and it can be violated in the classical
approach.

This inequality provide a lower bound for the quantum entropy of the
universe. We think that there can be a relation of this lower
quantum bound with the lower bounds for entropy discussed in
\cite{verlinde}.

We are going to discuss this aspect in a future work.


\begin{thebibliography}{99}
\bibitem{Manko:2003dp}
  V.~I.~Manko, G.~Marmo and C.~Stornaiolo,
  Gen.\ Rel.\ Grav.\  {\bf 37}, 99 (2005)
  [arXiv:gr-qc/0307084].

\bibitem{Man'ko:2004zj}
  V.~I.~Man'ko, G.~Marmo and C.~Stornaiolo,
`Cosmological dynamics in tomographic probability representation,''
to appear in Gen.\ Rel.\ Grav.\  {\bf 37} (2005)
  arXiv:gr-qc/0412091.

\bibitem{Stornaiolo:2006kz}
  C.~Stornaiolo,
  J.\ Phys.\ Conf.\ Ser.\  {\bf 33} (2006) 242.

\bibitem{Stornaiolo:2006da}
  C.~Stornaiolo,
  AIP Conf.\ Proc.\  {\bf 841} (2006) 645.

\bibitem{hawking1} S. W. Hawking, Nucl. Phys {\bf B239}, 257 (1984)
 \bibitem {whdw} B. S. DeWitt Phys. Rev.  {\bf 160}, 1113, (1983); J. A  Wheeler in
 \textit{Battelle Rencontres}, edited by C. DeWitt and J. A. Wheeler (Benjamin, New
 York, 1968).
\bibitem {hh} J. B. Hartle and S. W. Hawking, Phys. Rev. D {\bf 28}, 2960, (1983)

 \bibitem {page} D. N. Page Phys. Rev. D {\bf 34}, 2267, (1986)
 \bibitem {hawking2} S. W. Hawking Phys. Scr. {\bf T15}, 151
 (1987)
 \bibitem {wigner32} E. Wigner, Phys. Rev. {\bf 40}, 749 (1932)
 \bibitem {anton} F. Antonsen, ``Deformation Quantisation of
 Gravity'', gr-qc/9712012

 \bibitem {marmops} O. V.
Manko, V.I. Manko, G.~Marmo, Phys. Scr. {\bf62}, 446, (2000).
\bibitem {marmoolga} O. V. Manko, V. I. Manko, G.~Marmo,  J. Phys.
A {\bf 35}, 699, (2002)
 \bibitem {fronsdal} F. Bajen,
M. Flato, M. Fronsdal, C. Lichnerowicz, D.
 Sternheimer, Lett. Math. Phys. {\bf 1}, 521 (1975)


 \bibitem {marmo2} G. Esposito, G. Marmo, G. Sudarshan, ``From Classical to Quantum Mechanics
An Introduction to the Formalism, Foundations and Applications''
Cambridge University Press (2004).


\bibitem {mancini} S.Mancini, V.I.Manko, P.Tombesi, J.Opt.B: Quantum and Semiclass.
Opt., \textbf{ 7},  615 (1995)
 \bibitem {lecture}O. V. Manko, V. I. Manko, J. Russ. Laser Res. {\bf 18},
 407 (1997)
  \bibitem {mendespl} V. I. Manko, R.V. Mendes, Phys. Lett.
{\bf A263}, 53 (1999)
\bibitem {mendes2} V.I. Manko, R. V. Mendes,
Physica {\bf D145}, 330 (2000)

 
\bibitem{Stornaiolo:1994mw}
C.~Stornaiolo,
Phys.\ Lett.\ A {\bf 189} (1994) 351.

\bibitem{Lemos:1995qu}
N.~A.~Lemos,
J.\ Math.\ Phys.\  {\bf 37} (1996) 1449 [arXiv:gr-qc/9511082].
\bibitem{olga} O. V. Man'ko, Theor. Math. Phys. \textbf{ 121}, 285 (1999)
\bibitem{Faraoni:1999qu}
V.~Faraoni,
Am.\ J.\ Phys.\  {\bf 67} (1999) 732 [arXiv:physics/9901006].
\bibitem{Lapchinsky:1977vb}
V.~G.~Lapchinsky and V.~A.~Rubakov,
Teor.\ Mat.\ Fiz.\  {\bf 33} (1977) 364.



\bibitem{marg} M. A. Manko, J. Russ. Laser Res. {\bf 22},
 168 (2001)

 \bibitem{mendes3}M. A. Manko, V . I. Manko, R. V. Mendes J Phys A  \textbf{34}, 8321 (2001)


\bibitem {gousheh} S. S. Gousheh, H. R. Sepangi, Phys. Lett.
{A272}, 304 (2000)

\bibitem{margarita} M. A. Manko, J. Russ. Laser Res. {\bf 21},
 421 (2000)

\bibitem{rosapl} V. I. Manko, L. Rosa, P. Vitale
Phys. Lett B {\bf 439}, 328 (1998)

\bibitem {d'ariano}
G.~M.~D'Ariano, S.~Mancini, V.~I.~Manko and P.~Tombesi,
Quantum Semiclass. Opt {\bf 8}, 1017, (1996), quant-ph/9606034.



 
 



\bibitem{shannon}C.~E.~Shannon, "A mathematical theory of communication," Bell
System Technical Journal, vol. 27, pp. 379-423 and 623-656,
July and October, 1948.  
\bibitem{renyi} A.~Renyi, Probability Theory (North-Holland, Amsterdam, 1970)

  \bibitem{bialynicki} I~Bialynicki-Birula and J~Mycielski Comm. Math. Phys. {\bf 44}, 129,(1975)

\bibitem{183} V.~V.~Dodonov , V.~I.~Manko, Invariants and Evolution of Nonstationary Quantum Systems, Proceedings of the Lebedev Physical Institute, Nauka (Moscow) (1987) vol.183 [translated by Nova Science, Commack, New York (1989)]
\bibitem{Hartle:1997hw}
  J.~B.~Hartle,
 ``Quantum cosmology: Problems for the 21st century,in ``PHYSICS IN THE 21ST CENTURY'': Proceedings. Edited by K. Kikkawa, H. Kunitomo, H. Ohtsubo. Singapore, World Scientific, 1997.
  arXiv:gr-qc/9701022.
\bibitem{renato1}S.~De Nicola, R.~Fedele, M.~A.~Man'ko and V.~I.~Man'ko Eur. Phys. J. B 52, 191–198 (2006)

\bibitem{renato2} S.~De Nicola, R.~Fedele, M.~A.~Man'ko,
V~.I.~Man'ko ``New Inequalities for Tomograms in the Probability
Representation of Quantum States''  Acta Hungarica (2006) in press  quant-ph/0611114







\bibitem{Bekenstein:1980jp}
  J.~D.~Bekenstein,
   ``A Universal Upper Bound On The Entropy To Energy Ratio For Bounded
   Systems,''
  Phys.\ Rev.\ D {\bf 23} (1981) 287.
\bibitem{verlinde}
E.~P.~Verlinde,
  ``On the holographic principle in a radiation dominated universe,''
  arXiv:hep-th/0008140.





\end{thebibliography}
\end{document}